\documentclass[5p]{elsarticle} 

\usepackage{lineno,hyperref}
\modulolinenumbers[5]

\usepackage{amsmath,amssymb,amsfonts}
\usepackage{algorithmic}
\usepackage{graphicx}
\usepackage{textcomp}
\usepackage{xcolor}

\newcommand{\Qirr}{Q_{\text{SEI,irr}}} 
\newcommand{\DLiI}{D^{e^{\text{-}}}}
\newcommand{\cLiINull}{c^{{e^{\text{-}}}}_0}
\newcommand{\cLiI}{c^{e^{\text{-}}}}
\newcommand{\LiI}{\text{e}^\text{-}}

\newcommand{\Ds}{D_{\text{s}}}

\newcommand{\csmax}{c_{\text{s,max}}}
\newcommand{\Phis}{\Phi_{\text{s}}}
\newcommand{\phie}{\varphi_{\text{e}}}

\newcommand{\dSEI}{L_{\text{SEI}}}
\newcommand{\dSEIo}{L_{\text{SEI,0}}}
\newcommand{\VSEI}{V_{\text{SEI}}}
\newcommand{\sSEI}{s_{\text{SEI}}}
\newcommand{\NSEI}{N_{\text{SEI}}}
\newcommand{\NSEIdiff}{N_{\text{SEI,diff}}}

\newcommand{\USEI}{U_{\text{SEI}}}
\newcommand{\etaSEI}{\eta_{\text{SEI}}}
\newcommand{\kappaSEI}{\kappa_{\text{Li}^+}^{\text{SEI}}}

\newcommand{\jint}{j_{\text{Li}^+}}
\newcommand{\etaBV}{\eta_{\text{s}}}

\newcommand{\Vexp}{V_{\text{exp}}}
\newcommand{\Vsim}{V_{\text{sim}}}

\newcommand{\eqBV}{A.5}
\newcommand{\eqBVOverpotential}{A.6}

\newcommand{\eqOCVanode}{A.8}
\newcommand{\eqOCVcathode}{A.9}

\newcommand{\figCTimages}{B.1}
\newcommand{\figHistogram}{B.2}

\newcommand{\tabElectrolyte}{B.1}
\newcommand{\tabPtwoDparams}{B.2}
\newcommand{\tabThreeDparams}{B.3}
\newcommand{\tabDegradationParams}{B.4}

\newcommand\red[1]{{\color{black}#1}}

\begin{document}

\begin{frontmatter}

\title{Microstructure-Resolved Degradation Simulation of Lithium-Ion Batteries in Space Applications\\
}

\author[affil1,affil2]{Linda J. Bolay}
\author[affil1,affil2]{Tobias Schmitt}
\author[affil1,affil2]{Simon Hein}
\author[affil4]{Omar S. Mendoza-Hernandez}
\author[affil6]{Eiji Hosono}
\author[affil7]{Daisuke Asakura}
\author[affil7]{Koichi Kinoshita}
\author[affil7]{Hirofumi Matsuda}
\author[affil8]{Minoru Umeda}
\author[affil4,affil5]{Yoshitsugu Sone}
\author[affil1,affil2,affil3]{Arnulf Latz}
\author[affil1,affil2,affil3]{Birger Horstmann\corref{correspondingauthor}}
\cortext[correspondingauthor]{Corresponding author}
\ead{birger.horstmann@dlr.de}

\address[affil1]{Institute of Engineering Thermodynamics, German Aerospace Center (DLR), Pfaffenwaldring 38-40, 70569 Stuttgart, Germany}
\address[affil2]{Helmholtz Institute Ulm (HIU), Helmholtzstraße 11, 89081 Ulm, Germany}
\address[affil3]{Institute of Electrochemistry, University of Ulm, Albert-Einstein-Allee 47, 89081 Ulm, Germany}
\address[affil4]{Institute of Space and Astronautical Science, Japan Aerospace Exploration Agency (JAXA), 3-1-1 Yoshinodai, Chou-ku, Sagamihara, Kanagawa 252-5210, Japan}
\address[affil5]{The Graduate University of Advanced Studies (SOKENDAI), 3-1-1 Yoshinodai, Chou-ku, Sagamihara, Kanagawa 252-5210, Japan}
\address[affil6]{Global Zero Emission Research Center, National Institute of Advanced Industrial Science and Technology (AIST), 1-1-1 Umezono, Tsukuba, Ibaraki 305-8568, Japan}
\address[affil7]{Research Institute of Energy Conservation, National Institute of Advanced Industrial Science and Technology (AIST), 1-1-1 Umezono, Tsukuba, Ibaraki 305-8568, Japan}
\address[affil8]{Department of Materials Science and Technology, Nagaoka University of Technology, 1603-1 Kamitomioka, Nagaoka, Niigata 940-2188, Japan}

\begin{abstract}
In-orbit satellite REIMEI, developed by the Japan Aerospace Exploration Agency, has been relying on off-the-shelf Li-ion batteries since its launch in 2005. The performance and durability of Li-ion batteries is impacted by various degradation mechanisms, one of which is the growth of the solid-electrolyte interphase (SEI). In this article, we analyse the REIMEI battery and parameterize a full-cell model with electrochemical cycling data, computer tomography images, and capacity fading experiments using image processing and surrogate optimization. We integrate a recent model for SEI growth into a full-cell model and simulate the degradation of batteries during cycling. To validate our model, we use experimental and in-flight data of the satellite batteries. Our combination of SEI growth model and microstructure-resolved 3D simulation shows, for the first time, experimentally observed inhomogeneities in the SEI thickness throughout the negative electrode for the degraded cells. 
\end{abstract}

\begin{keyword} 
Li-ion battery\sep solid-electrolyte interphase\sep REIMEI satellite\sep degradation model\sep microstructure-resolved simulation\sep parameter identification
\end{keyword}

\end{frontmatter}


\section{Introduction}

Lithium-ion batteries are the technology of choice for various applications like laptops, smartphones and electric vehicles. Since the early 2000s they are considered applicable for space missions or more precisely for low earth orbit (LEO) missions. The requirements for batteries in a LEO mission are to be reliable for about 5 years with 30,000 cycles when charged for ca. 60 minutes and discharged for ca. 30 minutes. 
\cite{Wang2005,Wang2006,Marsh2001,Smart2004}.
Battery reliability is hampered by the harsh conditions they have to withstand. Already during lift-off they experience severe vibrations. In space they suffer from extreme temperatures, radiation, vacuum, and microgravity \cite{Uno2011, Sone2016}. Thus, battery satellites are quite costly (for a small scientific satellite: 4 million US dollar \cite{Saito2011}). 

Failure of battery satellites may abort the satellite mission and can endanger other missions through space debris. Therefore, determination and prediction of battery state-of-health during operation is necessary. This is only possible by remote control of the satellite system from a control station on earth. The antenna receives the telemetry data of the battery and transmits operation commands, while the satellite passes the ground station during a small time window \cite{Saito2011}.

Typically, the cells are tested extensively prior to the mission start. Cycling experiments are conducted over long periods of time, adjusted to the planned mission. Additionally, the cells undergo various abuse tests. This ensures safe operability during the whole mission. 

If the mission changes for unforeseen reasons, the predictions for the aging of the cells have to be adjusted. A long time might have passed, however, since the start of the mission. Access to cells of the same kind than those used in the mission might be difficult. For these reasons, it is essential to have physics-based models available, which can predict the performance and degradation of the cells for any given operation of the satellite.  
These models are based on short term experiments. This yields the additional benefit of shortening the time for tests prior to the mission start.

The major degradation process taking place in Li-ion batteries is the growth of the solid-electrolyte interphase (SEI). This passivation layer builds on the anode surface during the first cycles of the cell, when solvents from the electrolyte react with Li-ions and electrons. This process prevents the electrolyte from further reaction but does not passivate the reaction completely. The SEI is growing continuously during storage as well as during cycling. Thereby, Li-ions are consumed irreversibly, which causes capacity loss of the cell. 

The growth of the SEI is discussed extensively in literature, both in experiment and in theory. Usually, either short-term growth, which considers growth behavior during the first cycle, or long-term growth, which considers continuous growth over months and years, is investigated. In this work, only long-term degradation is discussed.

In experiments, a square-root-of-time ($\sqrt{t}$) behavior is observed for long-term growth \cite{Brown2008a, Uno2011}. 
Furthermore, experiments show that SEI growth is accelerated during cycling compared to storage \cite{Brown2008a}.
Attia et al. show that the SEI is especially growing faster during charging of the cell \cite{Attia2019}.
Another aspect is the inhomogeneous growth of the SEI, which is investigated on nanoscale \cite{Boniface2016, Huang2019, Rahe2019} and on cell level \cite{Pfrang2018}.

\red{Short-term SEI growth, i.e. the initial build-up of the SEI during the first cycles, is addressed by atomistic  theories like density functional theory and molecular dynamic simulations \cite{Wang2018,Bertolini2018}.
	The long-term growth behavior of the SEI can be described with empirical models \cite{Schmalstieg2014,Severson2019}. But a better understanding of the internal processes is given by physics-based models. For these, several transport mechanisms are considered as the rate-limiting process. The most important are electron conduction \cite{Roder2017,Das2019}, electron tunneling \cite{Lin2016,Li2015}, reaction kinetics \cite{Heinrich2019,Chouchane2021}, radical diffusion \cite{Single2018,Soto2015}, and solvent diffusion \cite{Pinson2013,Ploehn2004,Ekstrom2015,Roder2016}. 
	Single et al. compared all of these transport processes to storage experiments and show that the most probable mechanism for SEI growth during storage is radical diffusion \cite{Single2018}.
	Das et al. model the asymmetric growth behavior which was shown in Ref.~\cite{Attia2019} by coupling SEI growth to the Li-ion concentration in the SEI \cite{Das2019}.
	Von Kolzenberg et al. extended the model of radical diffusion in Ref.~\cite{Single2018} by an electron migration effect during cycling, which also reproduces the asymmetric growth. They describe the transitions between transport- and reaction-limited growth and with that give a consistent explanation for the short-term and long-term growth behavior \cite{VonKolzenberg2020}. 
	A comprehensive review of SEI models is given by Horstmann et al. in Ref.~\cite{Horstmann2019}. 
}

Until recently, electrode level continuum models had not reproduced the experimentally observed SEI heterogeneity \cite{Pinson2013,Tahmasbi_2017}. \red{In his PhD thesis, Schmitt implemented a preliminary version of the model of Ref.~\cite{VonKolzenberg2020} in the software BEST \cite{BEST} and simulated the growth behavior in a 3D resolved microstructure}. Thereby, it was shown that the SEI is growing inhomogeneously \cite{Schmitt2020}. This was the first time that experimentally observed inhomogeneities in the SEI thickness could be shown with a physics-based model. Considering reaction-limited growth, such heterogeneous SEI growth has also been recently analyzed by Chouchane et al. \cite{Chouchane2021}.

In this article, we discuss how to incorporate the model of von Kolzenberg et al. into the physics-based model for Li-ion batteries of Latz et al. \cite{Latz2011} and simulate the SEI growth in a pseudo-two-dimensional (P2D) framework and in 3D. \red{We are the first to show inhomogeneous long-term growth of the SEI in microstructural-resolved simulations.} To validate our model, we make use of experiments and in-flight data of satellite REIMEI's battery cells. 

REIMEI is a small scientific satellite developed by the Japanese Aerospace Exploration Agency (JAXA). It was launched in 2005 with the main scientific mission to observe the aurora and demonstrate advanced satellite technology, one of which is Li-ion batteries that use off-the-shelf pouch cells. In 2013 the aurora observation was terminated. The satellite operation was extended with the main focus shifted to the prediction of performance and lifetime of the on-board Li-ion batteries \cite{Sone2016,Mendoza2020}. 

For a profound analysis of the battery state, a physics-based model that describes the cell degradation is required. However, only little is known about the cells parameters, as a long time has passed since the start of the mission. This makes the modeling a challenging task.

\red{One part of this challenge is the parameterization of the P2D model. This is addressed by different optimization approaches, where the error between experiment and model simulation, depending on the parameters, is to be minimized. Different approaches can be found in the literature. Forman et al. use a genetic algorithm to identify the parameters \cite{Forman2012}. Rahman et al. use a particle swarm optimization algorithm \cite{Rahman2016}. Here, we extract some parameters from CT images by image processing. The remaining parameters are determined with a surrogate optimization. To the best of our knowledge, this optimization method has not yet been used for battery models.}

In sec.~\ref{sec:REIMEI}, we give a brief overview of REIMEI and its batteries. After that, we describe the SEI model in detail in sec.~\ref{sec:Model}, which we then use for simulating the cycling and the degradation of the cells. For this purpose, in sec.~\ref{sec:BatteryCycling}, we first determine the parameters of the fresh cell with an optimization approach. Subsequently, in sec.~\ref{sec:Discussion} we determine the parameters of the degradation model and discuss the results of the degradation and SEI growth simulation.


\section{Experimental Data of REIMEI Batteries}\label{sec:REIMEI}


Satellite REIMEI, which flies on a polar low Earth orbit (LEO), has an orbital period of 96 minutes. During the sunlight period, the batteries get charged via the solar panels, while they discharge during the eclipse period to power the loads \cite{Uno2011}. This results in 15 cycles a day.
To ensure a reliable operability of 3~-~5 years, the temperature of the batteries is maintained at $19-22^\circ$C \cite{Saito2011, Sone2016} and the depth of discharge around 20~\% \cite{Uno2011}.

Two identical batteries are connected in parallel on board of the satellite REIMEI, each consisting of seven pouch cells connected in series. The lithium-ion cells are produced by NEC-Tokin Corporation. The positive active material is spinel lithium manganese oxide (LMO) and the negative electrode consists of graphite. The liquid electrolyte is 1~M LiPF$_6$ in EC/DEC (3:7 by wt\%) with additives \cite{Sone2016, Uno2011}.
The cells have a voltage range of \mbox{3.0~V - 4.1~V (4.2~V)}. Their nominal capacity is 3~Ah.
\cite{Brown2008a}

The in-flight operation of the cells depend on the loads in use. 
Given by the on-board devices, the cells reach their end of life when the end-of-discharge voltage (EoDV) falls below 3.75~V \cite{Uno2011}.

For the parameterization of the models, two kinds of terrestrial measurements are used. The electric battery response was measured after the mission began (see sec.~\ref{sec:Electrochemistry}). We identify parameters of the cell microstructure from CT images (see sec.~\ref{sec:CT}). 

\subsection{Terrestrial electrochemical measurements}\label{sec:Electrochemistry}

Accompanying the REIMEI mission, Uno et al. performed terrestrial cycling experiments \cite{Uno2011} with the same kind of cells as those of the satellite. The experiments span about 27,000 cycles, which is equivalent to five years. They charged and discharged several cells with a cycling protocol similar to the one of the satellite. To track the degradation of the cell, they measured the remaining capacity and the EoDV of the cells. 

In the experiments, the cells get discharged with a constant current (CC) of 1~A for 35~minutes, which corresponds to approximately 20~\% DOD. Then they get charged for 65~minutes with a constant current - constant voltage (CC-CV) profile. The charging is performed at 1.5~A to the upper voltage limit of 4.1~V. After each discharging, the EoDV is measured. As the cells are degrading, the EoDV falls below 3.75~V after several thousand cycles. At that point, the upper voltage limit for the charging gets changed to 4.2~V to prolong the remaining useful life of the cells. Before and during the cycling, they measure the remaining capacity of the cells several times by discharging it with a CC of 1~A to 3.0~V.  We use these electrochemical measurements of Uno et al. in sec.~\ref{sec:Parameterization} to parameterize the P2D model. 
\red{For the model we assume constant temperature and constant pressure throughout the cell. This assumption is based on the experiments in Ref.~\cite{Uno2011} and on the in-flight data as can be seen in Ref.~\cite{Sone2017}.}
In sec.~\ref{sec:Parameterization:Degradation}, the measurements are used to parameterize the degradation.

In 2008, Brown et al. conducted comprehensive terrestrial experiments at JAXA.
They investigated the degradation of cycled and calendaric aged cells under various conditions. For the cycling, they use the same cycling protocol than Uno et al. and vary it regarding temperatures and discharge currents respectively the depth of discharge (DOD). They examined full and half cells before, during and after the aging process. Furthermore, they measured open-circuit voltage (OCV) curves for both electrodes \cite{Brown2008a, Brown2008b}.
The OCV curve measurements of Brown et al. are used in sec.~\ref{sec:Parameterization} for the P2D model. Furthermore, our cycling protocols in sec.~\ref{sec:3D} are based on those of Browns et al.

\subsection{Computer tomography}\label{sec:CT}

One of the pouch cells of the kind used in the satellite was stored by JAXA. Several years after the start of the mission, CT images of the cell, which was stored at room temperature, were taken in various resolutions. Examples of the CT images can be found in Fig.~\figCTimages \ in the supporting information.
From the CT images we extract several parameters of the cell structure. These are the thicknesses and porosities of the cell layers and the specific surface area of the electrodes. 
These values are necessary to parameterize the model described in sec.~\ref{sec:Model}. 

The thickness $L$ of the layers is determined using the CT images with 1~$\mu$m and 3.5~$\mu$m resolution. To separate the layers, we consider grayvalues of voxels in the three dimensional CT image. Every type of layer in the cell has a specific grayvalue distribution. We assign the median of grayvalues to every voxel layer parallel to the cell layers. Hereby, we get a one dimensional signal which is perpendicular to the cell layers. The changes of grayvalue medians are determined by signal processing methods. These represent the boundaries between cell layers. The thickness is then averaged over several layers. 

The porosities $\epsilon$ of the different electrodes are determined from a representative section of the 1 $\mu$m resolution CT image. The histogram of grayvalues within this volume is determined and approximated by three normal distributions. For the cathode, the histogram and the corresponding distribution approximation can be seen in Fig.~\figHistogram \ in the supporting information.
The integral of these distributions yield the volume fractions of the phases in the electrode, the pore space, the active solid phase, and the passive solid phase, which consists of binder and conductive material. From the distributions, we get a binarization of the electrode volume using k-means clustering. We use a simplified version of the method described in Ref.~\cite{Prifling2019}. Here, we only want to classify the voxels as pore space or solid phase. The classification is assumed sufficient, when the grayvalue distributions classes have the same integral than the approximated distributions. The binarization is then used to determine the specific surface area $A_{\text{spec}}$ of the electrodes. This is done with the software GeoDict \cite{geodict}. 

In the case of the separator, we proceed in the same way. However, here the histogram of the grayvalues is approximated by only two normal distributions. The porosity is then again obtained from the integral of the distributions. 

For the full cell area $A_{\text{cell}}$, the width, length, and number of cell layers are needed. The length is taken from the CT image with 140~$\mu$m resolution, which shows the whole cell. The width and number of layers are taken from the CT image with 24~$\mu$m resolution, which shows the upper part of the cell. We get $A_{\text{cell}} = 0.139 \ \mathrm{m^2}$. The other identified parameters can be found in Table~\tabPtwoDparams \ in the supporting information.


\section{Computational Model}\label{sec:Model}

In order to understand and simulate the processes taking place in the cell, a physics-based model for the short-term cycling as well as the long-term degradation behavior of the cells is necessary.
The long-term degradation of the cell is mainly determined by capacity fading, which is caused by the loss of Li-ion inventory. 
To quantify the long-term capacity fading under cycling as well as under storage conditions, a new physics-based model has been developed. Von Kolzenberg et al. describe the model in detail in Ref.~\cite{VonKolzenberg2020}. Here, we summarize it and embed it into the thermodynamic consistent transport theory for Li-ion batteries of Latz et al. \cite{Latz2011}.

\begin{figure}[t!]
	\centerline{
		\includegraphics[width=0.48\textwidth]{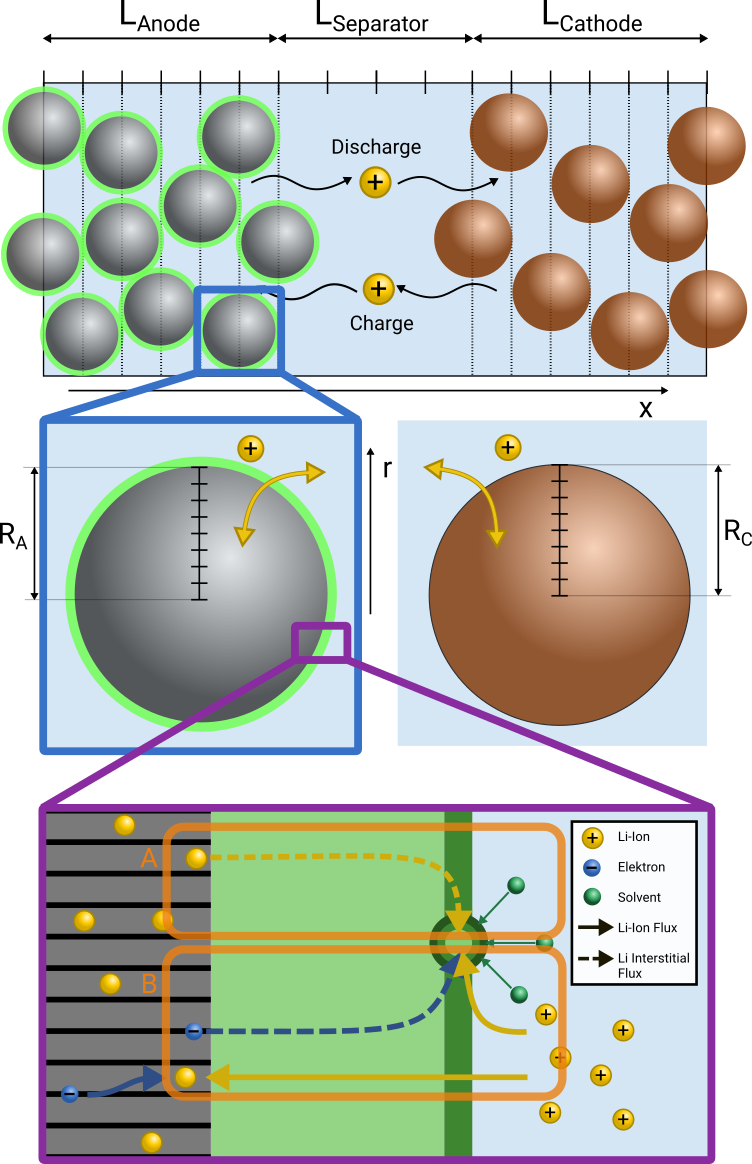}
		\put(-255,380){a)}
		\put(-255,240){b)}
		\put(-255,110){c)}
	}
	\caption{Lithium transport on different scales. a)~Li-ion transport in electrolyte between electrode particles. b)~Li-ion current at electrode interface. c)~SEI growth during calendar aging (A) and during cycling (A~+~B).}
	\label{fig:models}
\end{figure}

\red{This theory models transport and reactions of Li-ions in electrolyte and electrode particles on the microstructural scale of batteries. Several equations describe how the inner states, like Li-ion concentration or electrical potential, are distributed in the cell and how they evolve over time. The corresponding equations can be found in the supporting information \cite{Latz2011}.
	For the microstructure-resolved simulations, first, the morphology of the electrodes needs to be reconstructed from tomographic image data or simulated with given morphology characteristics. For example, this is done with stochastic geometry methods by Hein et al. \cite{Hein2016a}. In our research group, this 3D representation of the Li-ion cell is then used with the software BEST to perform spatially resolved electrochemical simulations \cite{BEST,Latz2015}. In Ref.~\cite{Hein2016} Hein et al. use these simulations to study and analyze the heterogeneous surface effects
	(see also sec.~\ref{sec:Discussion}).}

Using volume averaging techniques the transport theory by Latz et al. can be uspcaled into a pseudo-mul\-ti\-di\-men\-sion\-al model \cite{Schmitt2020a} like the one dimensional P2D model, which is illustrated in Fig 1. 
This model has first been described by Doyle et al. \cite{Doyle1993}. It has since been used a lot to describe Li-ion battery dynamics \cite{Bizeray2015,Forman2012,Rahman2016}.
In Fig.~\ref{fig:models}a the volume-averaged transport of Li-ions in the electrolyte between electrodes is shown. The interfacial reaction kinetics between electrolyte and electrode is depicted in Fig.~\ref{fig:models}b. It is modeled by a Butler-Volmer rate, with the overpotential $\etaBV$ dependent current density $\jint(\etaBV)$ (see eq.~\eqBV \ and \eqBVOverpotential \ in the supporting information).

When a Li-ion cell gets charged for the first time, the electrolyte gets reduced at the anode surface due to the low graphite potential. The products build a passivation layer, the solid-electrolyte interphase (SEI), which prevents the electrolyte from further reduction. However, the SEI continues growing because of a leakage of electrons from the anode through the SEI to the electrolyte, where it reacts with Li-ions and solvents. This reaction consumes Li-ions irreversibly, which is one of the main reasons for capacity fading. 

The relation between irreversible capacity loss $\Qirr $, SEI thickness $\dSEI$, and the growth causing flux $\NSEI$ is given as 
\begin{equation}\label{eq:capacityLoss}
	\partial_t \Qirr = \frac{AF\sSEI}{\VSEI} \cdot \partial_t \dSEI = AF \cdot \NSEI, 
\end{equation}
where $\NSEI \geq 0$, $A$ is the surface area of the anode, $\sSEI$ is the mean stoichiometric coefficient of Li in the SEI formation reaction, $\VSEI$ is the mean partial molar volume of the SEI, and $F$ is the Faraday constant. 

Single et al. provided evidence that electron diffusion through the SEI is causing calendaric aging \cite{Single2018}. These electrons diffuse through localized state, e.g., neutral Li-interstitials diffusion \cite{doi:10.1021/ja305366r}. The corresponding flux is deduced from Fick's law and is given as
\begin{equation}
	\NSEIdiff = \DLiI \cdot \frac{ \cLiI}{\dSEI}, 
\end{equation}
where 
$\DLiI$ is the diffusion coefficient and $\cLiI$ is the interstitial concentration at the anode-SEI interface, which is dependent on the electrochemical potential of the anode relative to Li. This process is depicted as process A in Fig.~\ref{fig:models}c.

The capacity loss increases when the cell is cycled. Attia et al. show that the SEI grows almost exclusively during lithiation of the anode \cite{Attia2019}. 
This phenomenon is also modeled by an electron flux, which is influenced by the Li-ion current through the SEI. This can be seen in Fig.~\ref{fig:models}c process~B. To take the influence of the Li-ion current into account, an electron migration flux is incorporated and the SEI flux is derived from the Nernst-Planck equation

\begin{equation}\label{eq:SEI_flux1}
	\NSEI = \DLiI \cdot \frac{ \cLiI }{\dSEI} \ - \ z F \kappa_\text{SEI}^{\LiI} \nabla \Phi, 
\end{equation}
with the valence $z=-1$, the conductivity of the localized electrons $\kappa_\text{SEI}^{\LiI}$, 
and the potential gradient  $ \nabla \Phi $ in the SEI caused by the Li-ion flux.
According to \cite{Single2018}, the electron concentration in the SEI is approximated by 
\begin{equation}
	\cLiI = \cLiINull \cdot \exp \left(- \frac{F}{RT} \etaSEI \right),
\end{equation}
with the interstitial concentration $\cLiINull$ at $0V$ anode potential, the universal gas constant $R$, and the temperature $T$. 
Furthermore, the potential gradient is approximated by 
\begin{equation}
	\nabla \Phi = \frac{-\USEI}{\dSEI},
\end{equation}
where the potential drop is defined as 
\begin{equation}
	\USEI =  \frac{\dSEI}{\kappaSEI} \cdot \jint,
\end{equation}
with the conductivity $\kappaSEI$ of Li-ions in the SEI.

With the help of the Nernst-Einstein equation, the conductivity $\kappa_\text{SEI}^{\LiI}$ is approximated by the diffusion coefficient 
\begin{equation}
		\DLiI = RT\kappa_\text{SEI}^{\LiI}/\cLiI \approx 2RT\kappa_\text{SEI}^{\LiI}/\cLiINull
\end{equation}
and eq.~\ref{eq:SEI_flux1} is rearranged to 
\begin{equation}\label{eq:SEI_flux2}
	\begin{split}
		\NSEI  \ = \ & \frac{\DLiI \cdot \cLiINull }{\dSEI}  \cdot \exp \left(- \frac{F}{RT} \etaSEI \right) \\ 
		& \qquad \qquad \cdot \left( 1- \frac{F}{2RT} \USEI \right).
	\end{split}
\end{equation}

The overpotential for the reaction is given as 
\begin{equation}
	\etaSEI = \Phis - \phie - \USEI,
\end{equation}
where $\Phis$ is the anode potential and $\phie$ is the electrochemical potential of the electrolyte.

The growth of the SEI not only affects the cells capacity but also its inner resistance. This is taken into account in the Butler-Volmer rate for the intercalation into graphite, where the overpotential is supplemented by the potential drop in the SEI and becomes
\begin{equation}
	\Tilde{\eta}_{\text{s}} = \etaBV - \USEI.
\end{equation}
The overpotential of the Butler-Volmer term is given in Ref.~\cite{Latz2011} (see also eq.~\eqBVOverpotential \ in the supporting information).


\section{Battery Cycling}\label{sec:BatteryCycling}

We simulate battery cycling with the volume-averaged P2D model. Several battery parameters are required for this. The parameters describing the microstructure morphologies are extracted from CT data (see sec.~\ref{sec:CT}). The concentration-dependent electrolyte parameters are taken from Ref.~\cite{Ehrl2017}. The OCV curves are fitted to the measured curves in Ref.~\cite{Brown2008b} (see eq.~\eqOCVanode \ and \eqOCVcathode \ in the supporting information). The remaining cell parameters are determined with an optimization algorithm, where experimental data of a fresh cell from Ref.~\cite{Uno2011} are used. The parameters can be found in Table~\tabElectrolyte \ and \tabPtwoDparams \ in the supporting information.

\subsection{Parameterization}\label{sec:Parameterization}

The parameters of the cell model, that are not known from literature or measurements, are determined by an optimization, where the deviation between experiment and simulation, given as an objective function, is minimized. 

Those parameters are the initial SOC on the anode side and for both electrodes the maximal concentration of Li-ions in the active material~$\csmax$, the diffusion coefficient of Li-ions in the active material~$\Ds$, and the rate constant~$k$. 

The optimization algorithm should find the global minimum and preferably it is parallelizable as the evaluation of the time-evolution of our physics-based model is time-consuming. To this aim, Forman et al. \cite{Forman2012} used a genetic algorithm to identify the parameters of the P2D model and Rahman et al. \cite{Rahman2016} used particle swarm optimization.

Here, we will use the surrogate optimization, which is a stochastic optimization algorithm. The benefits of this algorithm are that its convergence to the global optimum is proven and that it can run in parallel \cite{Regis2007}. To the best of our knowledge, this is the first time that the surrogate optimization has been used to identify the parameters of the P2D model.

\subsubsection{Surrogate optimization}

We briefly outline the algorithm for surrogate optimization. For further details, we refer to Refs.~\cite{Regis2007,Wang2014}.
In a first phase, the algorithm evaluates the objective function for a given number of points, which are randomly chosen in the given parameter space. These points get interpolated by radial basis functions to build a surrogate function. In the next phase, the point with the smallest value is found. Around this point, multiple points are sampled and an additional merit function is evaluated at these points. 
The merit function consists of the surrogate and a function that describes the distance of sampled points to already evaluated points. 
This ensures that the surrogate function is minimized while searching in new regions. 
At the point, where the merit function has its minimal value, the objective function is evaluated and the surrogate function is updated. This procedure is repeated until a stopping criterion is reached. Subsequently, the surrogate gets rejected and a new surrogate is created with new random points, while the best fit of the previous surrogate is kept for comparison. This overall procedure of generating new surrogate functions can be repeated until the global optimum is found. Usually the search for the optimum is stopped after a predefined time.

\subsubsection{Fitting}\label{sec:Parameterization:Fitting}

The objective function for the optimization is a weight\-ed sum of the discharge curve error err$_\text{V}$ and the EoDV error err$_\text{EoDV}$. Here, we chose the L$^1$-Norm to measure the discharge curve error: 
\begin{equation}
	\text{err}_\text{V}(\theta) = \int_{T} | \Vexp (t) - \Vsim(t;\theta) | dt,
\end{equation}
where $\Vexp$ is the experimental voltage and $\Vsim$ is the voltage output of the simulation over the discharge time T for a given parameter vector $\theta$. 
The EoDV error gets measured after a few cycles, since the cycling range has to level off after full discharge in the experiment as well as in the simulation. 

The parameters are optimized in a reasonable parameter space. The results of the fitting can be found in Table~\tabPtwoDparams \ in the supporting information. As the algorithm stopped due to a time stopping criterion, it cannot be assumed that the resulting parameters are the optimal parameters. Also, it has not been shown yet that the parameters of the P2D model are identifiable at all. Bizeray et al. showed the identifiability for the single particle model, which is a reduced version of the P2D model, in Ref.~\cite{Bizeray2018}. In Ref.~\cite{Li2020}, Li et al. analysed the sensitivity of the P2D model parameters. They showed that only part of the parameters is identifiable. 

The parameters of the battery cycling and those of the degradation are correlated. Thus, the parameterization of both models needs to be linked to each other. For that purpose, the degradation parameters are taken into account in this section.


\begin{table}[b]
	\begin{center}
		\resizebox{0.48\textwidth}{!}{ 
			\begingroup
			\renewcommand{\arraystretch}{1.5} 
			\begin{tabular}{| c | c | c | c |} 
				\hline
				Protocol & Charge (CC-CV) & Discharge (CC) & Temperature \\
				\hline
				\hline
				P1 & 1.5 A / 4.1 V & 0.88 A \quad 0.74 A  &  20°C \\
				(in-flight) & 63 min & 15 min \quad 19 min & \\
				\hline
				P2 & 1.5 A / 4.1 V (4.2 V) &  1.0 A   &  25°C  \\
				(P2D) & 65 min & 35 min  & \\
				\hline
				P3 & 1.5 A / 4.1 V &  1.0 A   &  25°C  \\
				(3D) & 65 min & 35 min  & \\
				\hline
				P4 & 1.5 A / 4.1 V &  2.0 A   &  25°C  \\
				(3D) & 65 min & 35 min  & \\
				\hline
				P5 & 3.0 A / 4.1 V &  2.0 A   &  25°C  \\
				(3D) & 65 min & 35 min  & \\
				\hline
				
			\end{tabular}
		\endgroup
	}
	\caption[]{Cycling protocols used in experiments and simulations.}
	\label{table:CyclingProtocols}
\end{center}
\end{table}

\subsection{Simulation and validation with in-flight data}

\begin{figure}[t!]
	\centerline{\includegraphics[width=0.45\textwidth]{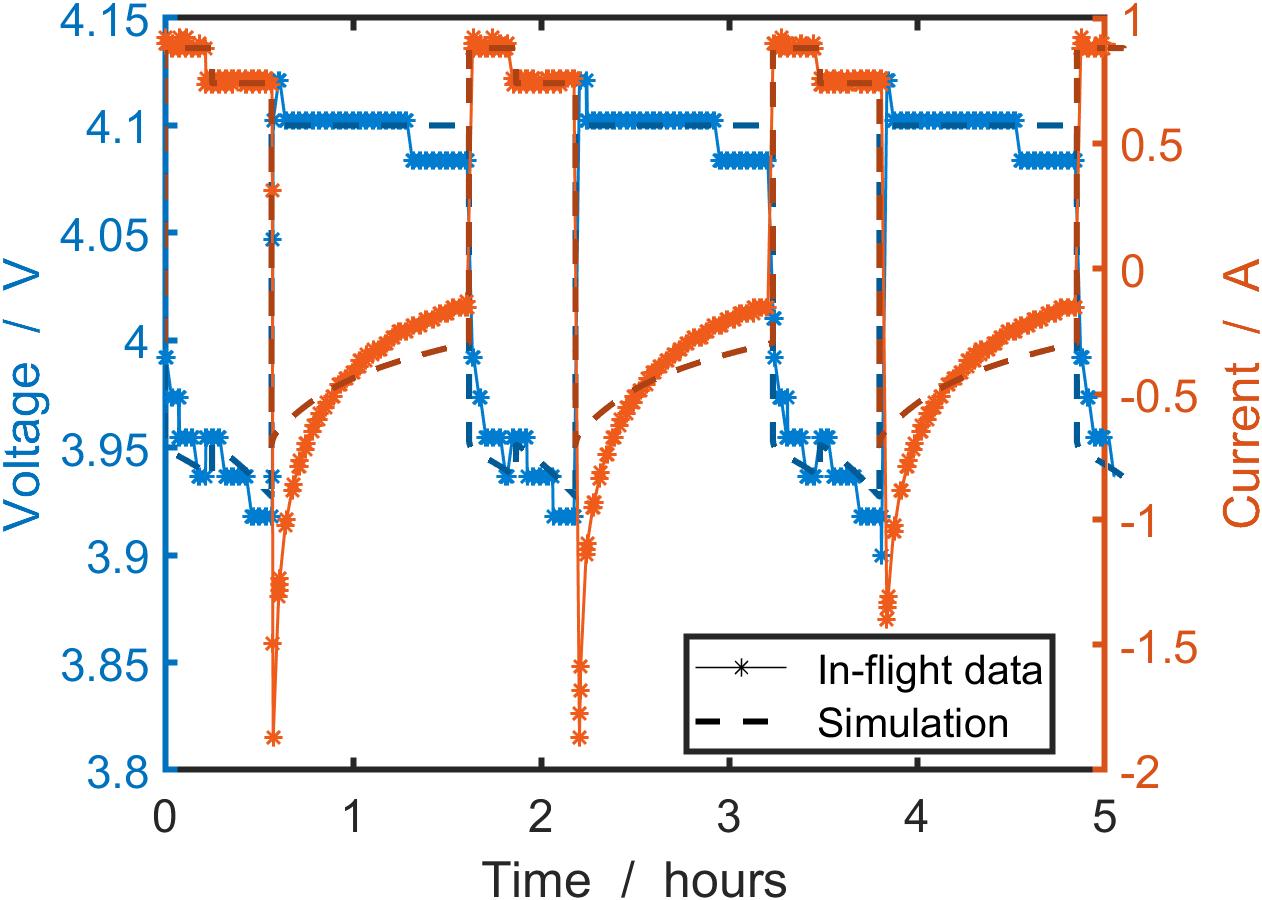}}
\caption{Three cycles of satellite REIMEI's battery. In-flight data and simulation corresponding to the satellite cycling protocol.}
\label{fig:SimInFlight}
\end{figure}

To validate the determined parameters, the satellite in-flight data are used. 
On orbit, the cells get charged via the solar panels with a CC-CV profile.
During the eclipse period, they get discharged to power the cameras for the aurora observation.
There are two different discharge profiles, depending on whether the aurora of the north or the south pole is observed. The cycling protocols are described in detail in Ref.~\cite{Brown2008a}. 

In Fig.~\ref{fig:SimInFlight} the simulation of the in-flight data is depicted. The data were taken from the early life of the battery in 2005. The cycling protocol, that has been used during this period of satellite operation, was for north pole observation. It can also be found in Table~\ref{table:CyclingProtocols} (P1).
\red{In the figure we show the agreement between simulation and experiment over the cause of three cycles. It can be seen that the discharge voltage curve is met by the simulation.}


\section{Battery Degradation}\label{sec:Discussion}
To determine the parameters of the degradation model, long-term cycling is simulated with the volume-averaged P2D model (see sec. \ref{sec:Parameterization:Degradation}). Subsequently, 3D simulations are performed with the determined parameters. Different cycling protocols are used to examine the influence of charge and discharge currents on the SEI growth in the microstructure (see sec. \ref{sec:3D}). 

The degradation with the P2D model is implemented in MATLAB \cite{MATLAB}. The 3D model is implemented with the software BEST \cite{BEST}. Both kinds of simulations are run on the high performance computer cluster JUSTUS~2 which belongs to the bwHPC.  

\subsection{Parameterization with P2D model}\label{sec:Parameterization:Degradation}

\begin{figure}[!t]
	\centering
		\includegraphics[width=0.41\textwidth]{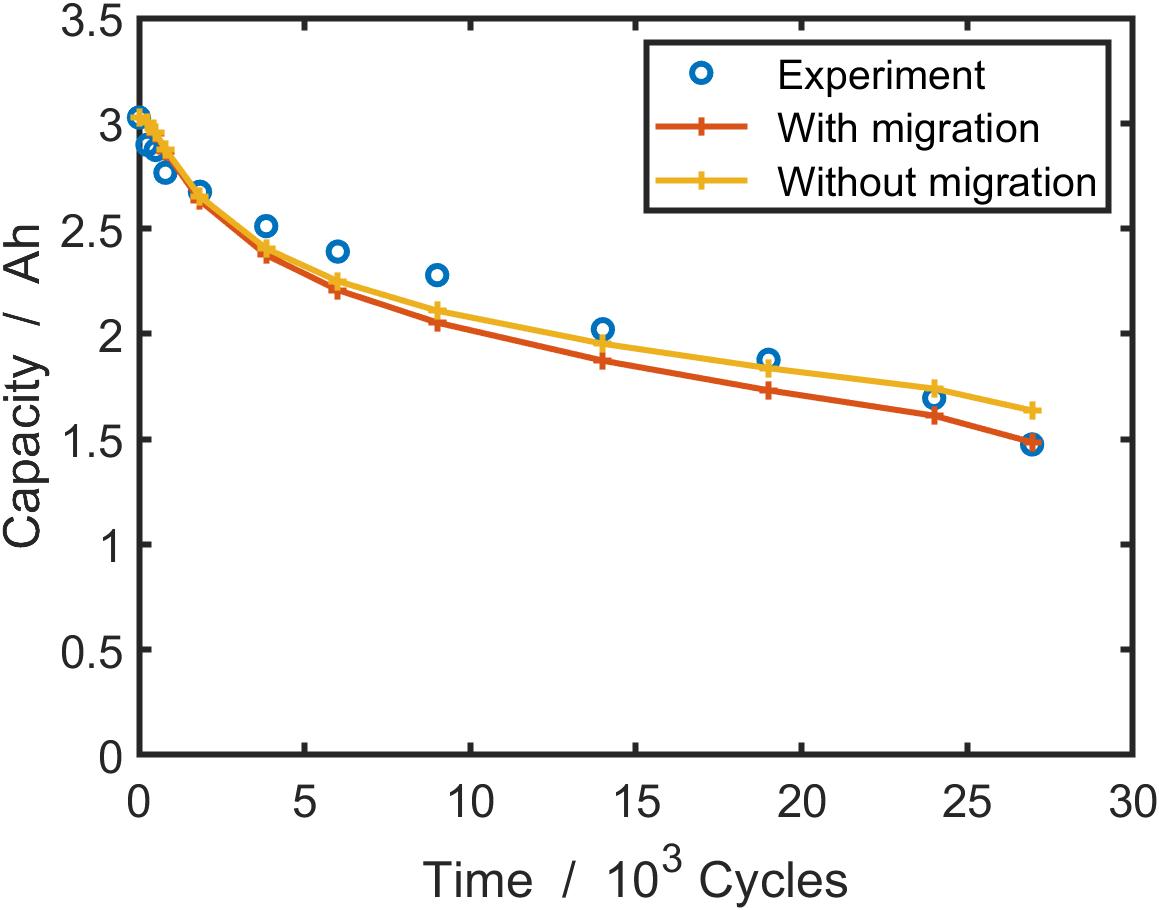}
		\put(-240,160){a)}
	
		\includegraphics[width=0.41\textwidth]{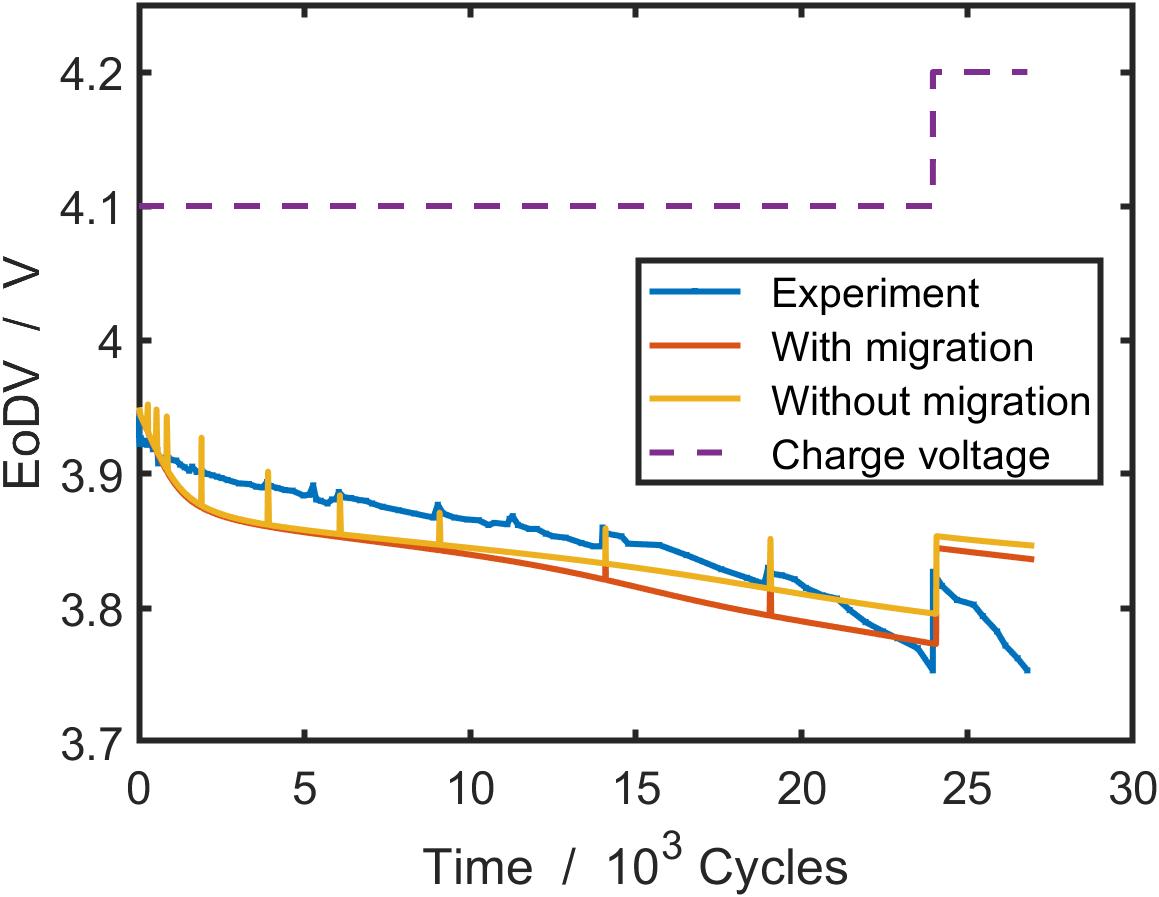} 
		\put(-240,160){b)}
	\caption{Two simulations of degradation experiment with satellite cycling protocol. Both simulations with equal parameters, one with and the other without electron migration flux. a)~Capacity and b)~end of discharge voltage (EoDV) of simulation and experiment.}
	\label{fig:1DCapacityEoDV}
\end{figure}

We simulate the long-term cycling protocol according to the experiments of Uno et al. \cite{Uno2011}. They cycled the cells about 27,000 times. After around 24,000 cycles, when the EoDV reaches 3.75~V, the charge voltage is changed from 4.1~V to 4.2~V. At 12 different points in time, the cells are fully discharged to 3.0~V to measure the remaining capacity. This can also be seen in Fig.~\ref{fig:1DCapacityEoDV}. The charge-discharge protocol used here can be found in Table~\ref{table:CyclingProtocols} (P2). 

The parameters, that are examined in this subsection, are the initial SEI thickness $\dSEIo$, the diffusion coefficient $\DLiI$ of localized electrons in the SEI, and the conductivity $\kappaSEI$ of Li-ions  in the SEI.
The other parameters, like the mean partial molar volume of the SEI, contribute to the model as multiplicative factors and hence are not identifiable.
Those parameters are taken from literature. All parameters of the degradation model are given in Table~\tabDegradationParams \ in the supporting information. 

The determination of the optimal parameter set is a multi-objective optimization problem, where the capacity fading as well as the EoDV error have to be minimized. 
As the simulation of around 27,000 cycles is quite expensive in terms of computation time, a sophisticated optimization algorithm is inappropriate. 
For that reason, a fixed parameter space is used, where the parameters are uniformly distributed in a reasonable range. 
The cycling is then simulated for all combinations of parameters simultaneously on the computer cluster. The optimal parameter set of the fixed parameter space is chosen.

In Fig.~\ref{fig:1DCapacityEoDV}a the capacity fading over around 27,000 cycles is depicted. Fig.~\ref{fig:1DCapacityEoDV}b shows the corresponding EoDV. At around 24,000 cycles, when the charge voltage is raised from 4.1~V to 4.2~V, the degradation accelerates. 
For the simulation, we first consider the red lines. In both graphs, we see that the degradation gets overrated at the beginning of the cycling. This can be explained by a not optimal choice of cell parameters. As we already described in sec.~\ref{sec:Parameterization:Fitting}, the optimization of the parameters is a difficult task due to computation time and the identifiability of the parameters. 
After around 2,000 cycles, over a wide range of the cycling, the trend of the simulation corresponds quite well to that of the experiment, especially in the EoDV. 
At the end, after about 20,000 cycles, the simulation underrates the degradation of the experiment. At this late period of cycling, there are probably additional degradation processes taking place in the cell, that are not covered by our SEI growth model. 
\red{This could be for example the cracking of anode particles as described in Ref.~\cite{Reniers2019} or Li plating. These processes lead, among other things, to a sudden increase of anode surface, which accelerates the SEI growth. 
	Mendoza-Hernandez et al. \cite{Mendoza2020} observe indicators for plating in the satellite in-flight data. These occur more frequently at a later stage of the cycling. Plating is responsible for or accelerates a number of degradation processes. This may be another reason why accelerated aging is observed that cannot be explained by SEI growth.}

\red{The difference between the new model of von Kolzenberg et al. \cite{VonKolzenberg2020} and the model of Single et al. \cite{Single2018} is the additional consideration of the Li-ion flux through the SEI during charging. This causes the migration of electrons due to an electric potential drop in addition to diffusion. To discuss the influence of the electron migration flux for SEI growth, we perform simulations of the degradation with and without the term that describes the electron migration ($ z F \kappa_\text{SEI}^{\LiI} \nabla \Phi = 0$) in eq.~\ref{eq:SEI_flux1}}. 
The simulation result is depicted in Fig.~\ref{fig:1DCapacityEoDV} as the yellow lines. It can be seen that the degradation decreases significantly in that case already after a few thousand cycles. Also the trend of the simulation does not correspond to the experiments in both graphs.

\subsection{3D microstructure-resolved simulations}\label{sec:3D}

\begin{figure*}[!t]
	\centerline{
		\includegraphics[width=0.85\textwidth]{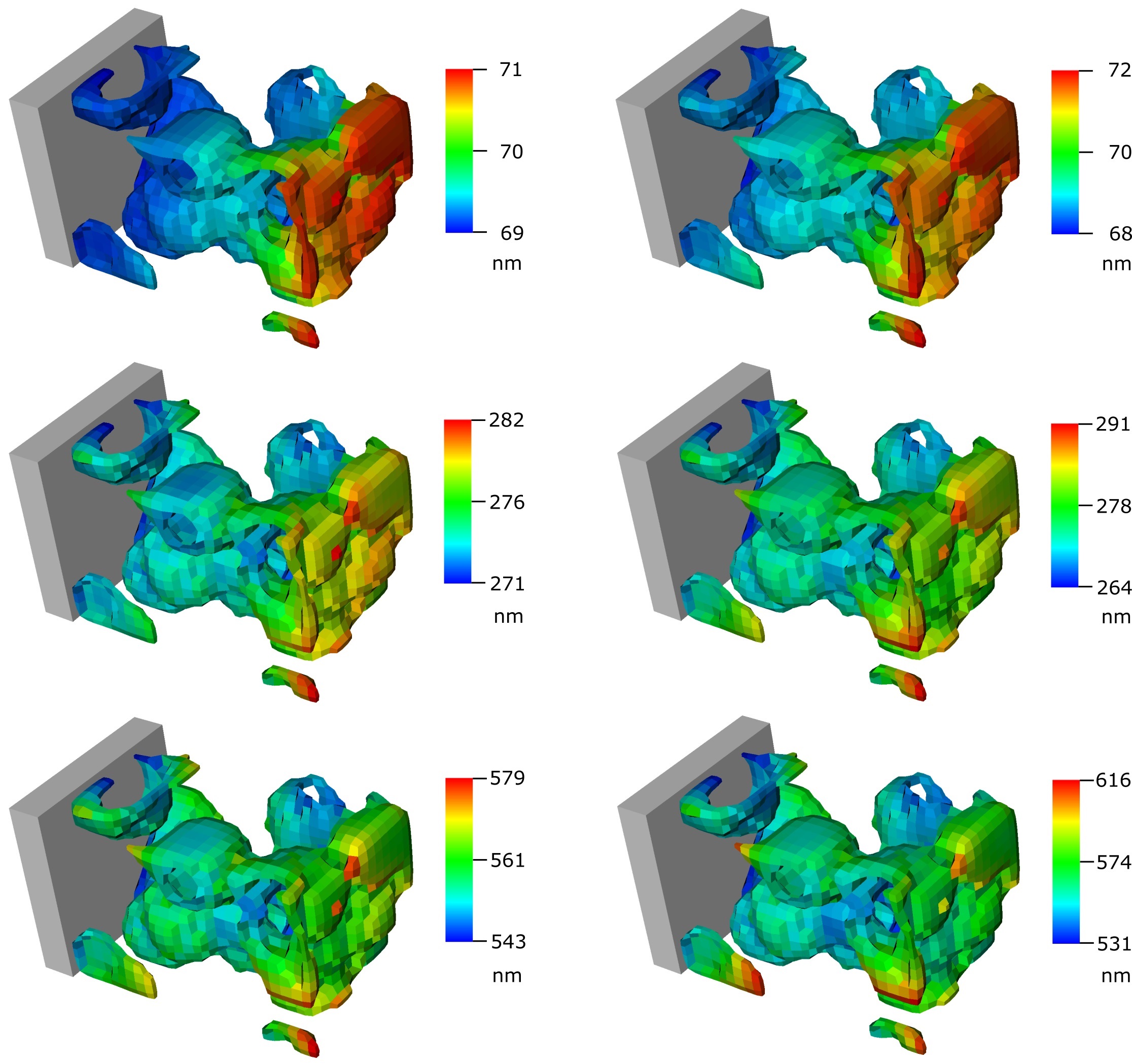}
		\put(-445,410){a)}
		\put(-445,270){b)}
		\put(-445,130){c)}
		\put(-205,410){d)}
		\put(-205,270){e)}
		\put(-205,130){f)}
	}
	\caption{SEI thickness of anode particles connected to current collector for two cycling protocols at three time points. Left column (a, b, and c) shows cycling protocol P3, right column (d, e and f) shows P4 (cf. Table~\ref{table:CyclingProtocols}). a) and d) 8 cycles. b) and e) 136 cycles. c) and f) 672 cycles.}
	\label{fig:SEIthicknessMicrostruct}
\end{figure*}
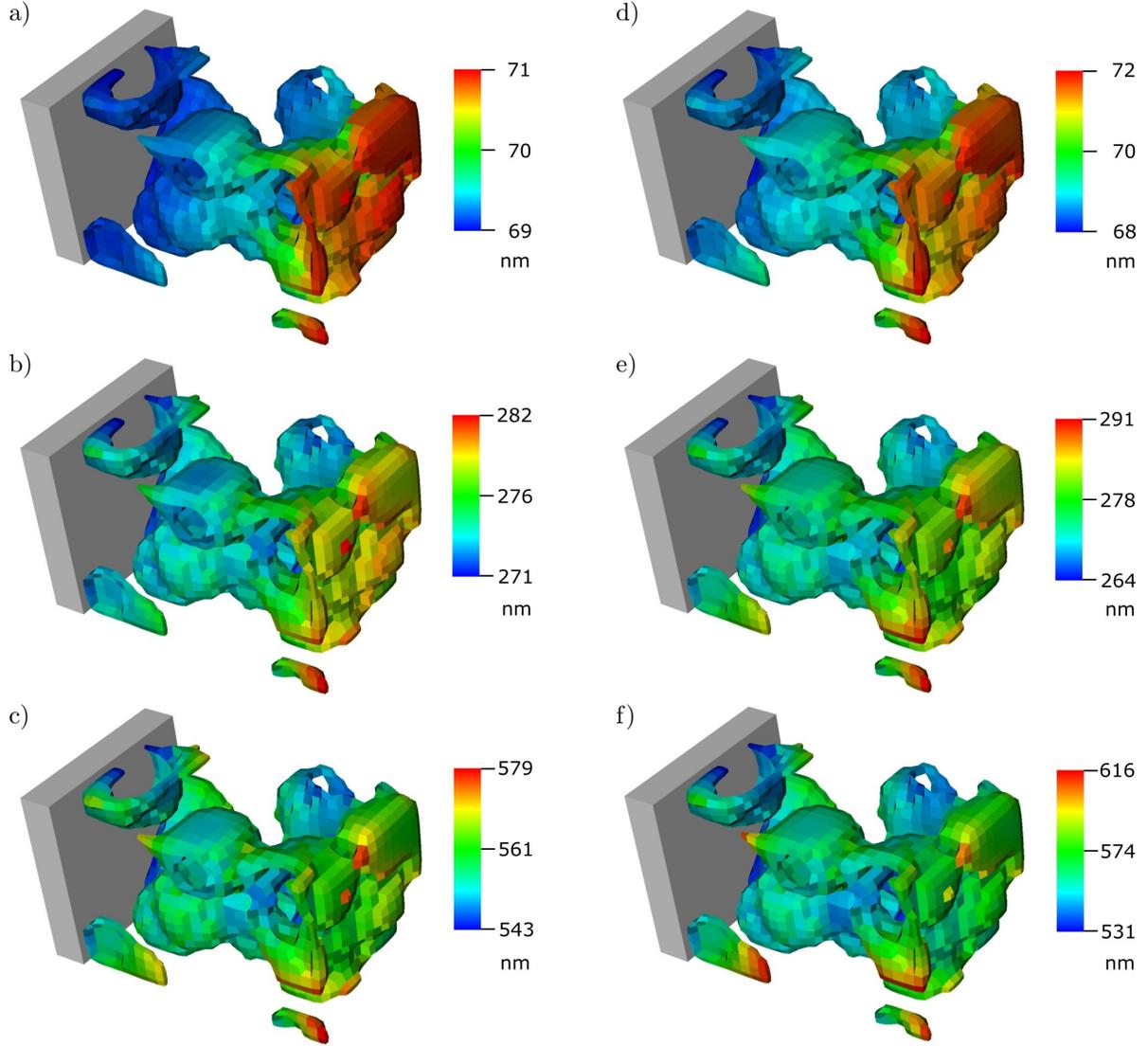

The microstructure-resolved SEI growth is simulated in 3D with the same degradation parameters as determined before with the P2D model. The cell and degradation parameters for the 3D simulations can be found in Table~\tabThreeDparams \ and \tabDegradationParams \ in the supporting information. There are three different cycling protocols used for the 3D simulations. They differ in charge and discharge currents to investigate how the current influences the SEI growth behavior. The protocols (P3~-~P5) can be found in Table~\ref{table:CyclingProtocols}. P3 is equivalent to the cycling protocol used with the P2D model (P2). In P4 the discharge current is doubled. This is based on the experiments in Ref.~\cite{Brown2008a}. In P5, both the charge current and the discharge current are twice as large as in P3. The intervals for charging and discharging are the same in all protocols and correspond to the typical satellite battery cycling protocols.

The 3D simulations reveal an inhomogeneous growth of the SEI. 
In Fig.~\ref{fig:SEIthicknessMicrostruct} the results of the simulations with two different cycling protocols at three points in time are shown. We see anode particles connected to a current collector. The color represents the thickness of the SEI after a certain amount of cycles. In the left column the cell is cycled with protocol P3, the right one is cycled with P4. Fig.~\ref{fig:SEIthicknessMicrostruct}a and d show the SEI thickness after 8 cycles, Fig.~\ref{fig:SEIthicknessMicrostruct}b and e after 136 cycles, and Fig.~\ref{fig:SEIthicknessMicrostruct}c and f after 672 cycles.  

For all states shown in Fig.~\ref{fig:SEIthicknessMicrostruct}, there are coexisting regions with a thicker SEI and regions with a thinner SEI.
In Fig.~\ref{fig:SEIthicknessMicrostruct}a and d we see that, at the beginning of the cycling, the particles near the separator have a thicker SEI than those near the current collector. Fig.~\ref{fig:SEIthicknessMicrostruct}b, c, e, and f show that after several cycles, the inhomogeneity spreads over the whole anode. When comparing Fig.~\ref{fig:SEIthicknessMicrostruct}b and e or Fig.~\ref{fig:SEIthicknessMicrostruct}c and f, we see that this happens sooner for the protocol with the higher discharge current. 

After many cycles, there are particular domains of the anode, where the SEI is noticeably thicker. In Fig.~\ref{fig:SEIthicknessMicrostruct}f, we observe that these are the thinner parts of the anode. Especially the peak-like structures have the thickest SEI. 

The inhomogeneous growth is caused by the varying SOCs through the anode and consequently varying overpotentials at the anode SEI interface. The thinner parts of the particles reach higher SOCs faster, when charging the cell. 


\begin{figure*}[htb!]
	\centering
	\begin{minipage}{0.5\linewidth}
		\centering
		\includegraphics[width=0.8\textwidth]{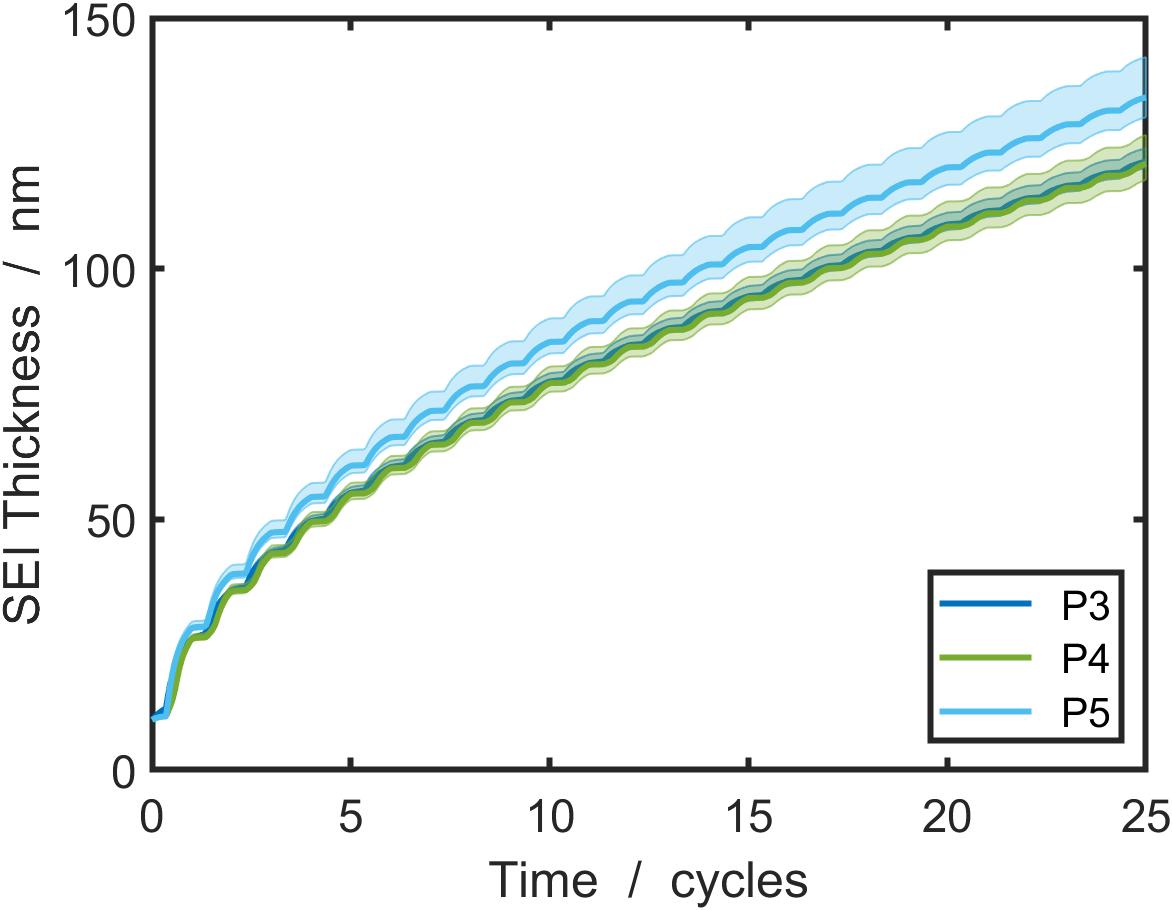} 
		\put(-240,160){a)}
	\end{minipage}%
	\begin{minipage}{0.5\linewidth}
		\centering
		\includegraphics[width=0.86\textwidth]{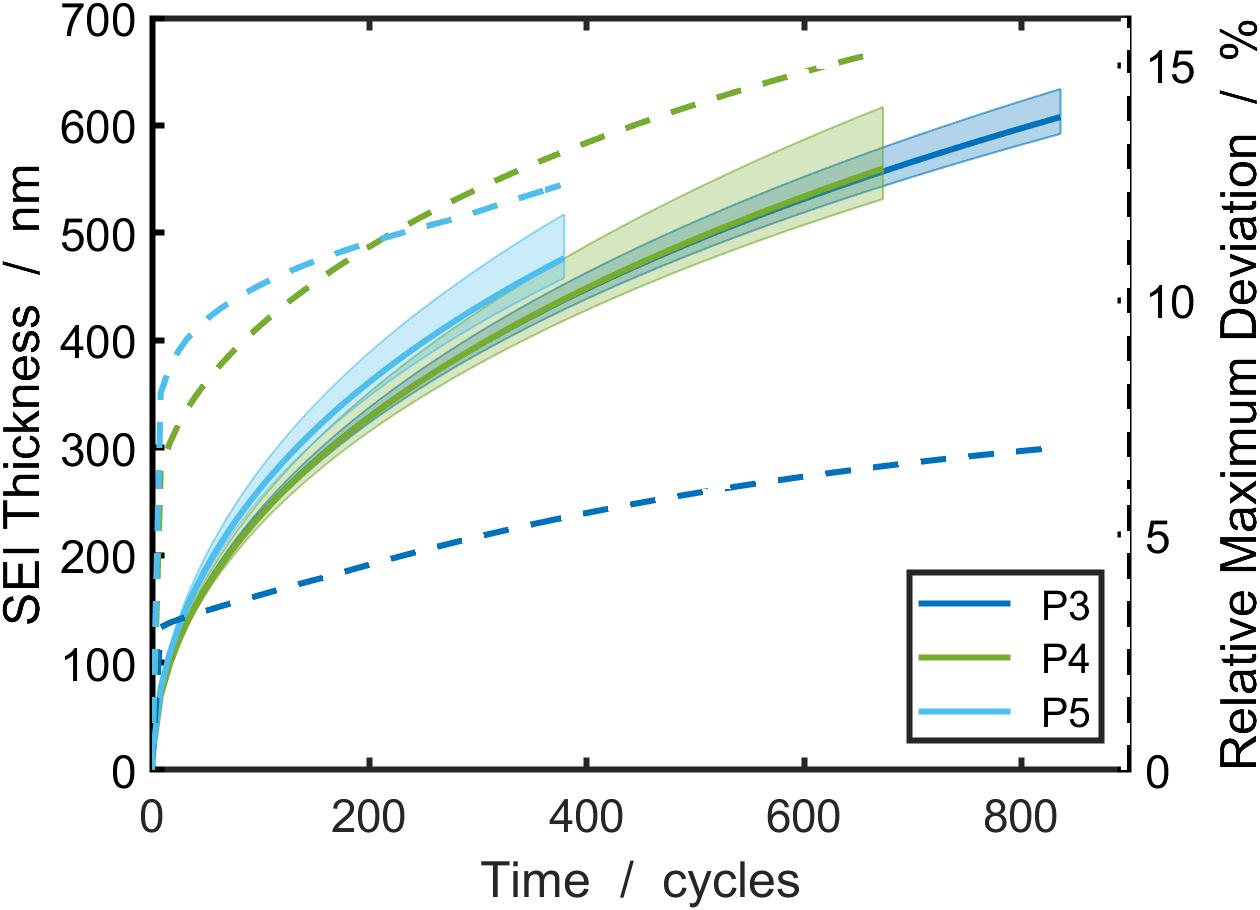} 
		\put(-240,160){b)}
	\end{minipage}
	\caption{Variation of SEI thickness in 3D simulation for three different cycling protocols (cf.~Table~\ref{table:CyclingProtocols}). a)~Stepwise growth behavior in each cycle. b)~SEI thickness and relative maximum deviation (dashed lines) over a long cycling period.}
	\label{fig:SEIthickness3D}
\end{figure*}

For a quantitative analysis of the growth and variation of SEI thickness, its mean, minimum, and maximum, throughout the anode, at every computed time step, is depicted in Fig.~\ref{fig:SEIthickness3D}. We investigate here the three cycling protocols P3~-~P5 at a short period of 25 cycles, which corresponds to about two days, and a long period of 400~-~800 cycles, which corresponds to approximately 1~-~2 months. 

When regarding the SEI thickness at the short period, as shown in Fig.~\ref{fig:SEIthickness3D}a, we observe a stepwise growth behavior. During charging of the cell or more precisely during lithiation of the anode, the SEI is growing faster than during delithiation. These simulations reflect that the growth rate depends on the current density, which has been observed experimentally in Ref.~\cite{Attia2019}.

Furthermore, when comparing the cycling protocols in Fig.~\ref{fig:SEIthickness3D}a, we note differences in SEI thickness and its inhomogeneity.
This becomes even clearer in the longer period in Fig.~\ref{fig:SEIthickness3D}b. The differences in the total length of the cycling simulations are due to numerical instabilities, when using higher currents. When comparing the mean SEI thickness of the different protocols, we observe that for P3 and P4, which differ in the discharge current, there is no significant difference in the thickness. In contrast, the SEI is growing faster for P5, which has a higher charge current than the other two protocols. This is also in accordance with the growth behavior observed in Ref.~\cite{Attia2019} and rationalized in Ref.~\cite{VonKolzenberg2020}. 

The relative maximum deviation, i.e. the difference between minimum and maximum divided by the mean of the thickness, is shown as dashed lines in Fig.~\ref{fig:SEIthickness3D}b. The simulations with doubled currents (P4 and P5) show a significantly larger deviation, i.e. a more inhomogeneous SEI growth. We also see that the deviation is already remarkable for the first few cycles and increases only slightly thereafter.


\section{Conclusion}

In this work, we present models for the short-term cycling and the long-term degradation behavior of a Li-ion cell. The modeling is based on experimental work and in-flight data of the batteries of small scientific satellite REIMEI, which was developed by JAXA. To describe the cycling behavior of the cell, we determine the parameters of the thermodynamic consistent transport theory of Latz et al. \cite{Latz2011}, which we incorporate in a P2D model. The parameters are determined with a surrogate optimization, in order to fit cycling experiments of the batteries. Subsequently, the model is validated by the satellite in-flight data.
\red{We could show that the surrogate optimization is appropriate to find parameters to fit simulation and experiments of a few cycles. But this method has the same limitation as other optimization methods, which lies in the unidentifiability of a few parameters of the P2D model.}
\red{One could use the surrogate optimization to fit a wider range of the experiment, e.g., the first 1000 cycles. This should improve the accuracy of the estimated parameters, and all degradation parameters could also be included in the optimization process. However, this would need significantly longer computation times.}

Furthermore, we give an overview of a model for the cell degradation, caused by SEI growth during storage as well as during cycling. This model describes the increased growth rate during lithiation of the anode. We determine the unknown parameters of the model with the help of long-term satellite battery experiments, where we focus on capacity fading and end-of-discharge voltage of the cells. 

With the same model and parameters, we simulate cell degradation in 3D. The simulations show an inhomogeneous growth of the SEI, which depends on the cycling currents. This provides a deeper understanding of the processes, which take place in a Li-ion battery. 

Apart from the capacity fading due to SEI growth, there are more degradation mechanisms, which have not been considered in this work. A degradation mechanism, which can cause severe damages in the cell and also contributes to capacity fading, is the deposition of metallic lithium on the anode surface. It is referred to as Li plating. In Ref.~\cite{Mendoza2020}, Mendoza et al. analyse the battery in-flight data of satellite REIMEI over an interval of 14 years. They observe a knee in the discharge voltage curves, which they suggest to be due to plating. Hein et al. modeled Li plating and simulated it in a microstructure-resolved framework \cite{Hein2016}. 

\red{In this study we chose physics-based continuum models for our simulations as these are able to reproduce the processes taking place in Li-ion batteries and make them more comprehensible. Apart from that, there are a lot more approaches to model the battery behavior. In Ref.~\cite{Shi2015} Shi et al. give a comprehensive overview on multi-scale computation methods. Similarly, Wang et al. review multi-scale models of SEI formation \cite{Wang2018}. To parameterize our continuum models, we employ experiments, since atomistic models do not yet provide the necessary accuracy.}

In future work, we will incorporate plating into our model. This work can be the foundation for performing state estimation and reliably predicting the state of health of satellite batteries in orbit.


\section*{Acknowledgments} 
This work was supported by the German Aerospace Center (DLR). 
The authors acknowledge support by the state of Baden-Württemberg through bwHPC and the German Research Foundation (DFG) through grant no INST 40/575-1 FUGG (JUSTUS 2 cluster). This work contributes to the research performed at CELEST (Center for Electrochemical Energy Storage Ulm-Karlsruhe).

\section*{Declaration of competing interest} 
The authors declare to have no competing interests.

\bibliography{DegradationModel}

\begin{thebibliography}{10}
\expandafter\ifx\csname url\endcsname\relax
  \def\url#1{\texttt{#1}}\fi
\expandafter\ifx\csname urlprefix\endcsname\relax\def\urlprefix{URL }\fi
\expandafter\ifx\csname href\endcsname\relax
  \def\href#1#2{#2} \def\path#1{#1}\fi

\bibitem{Wang2005}
X.~Wang, C.~Yamada, H.~Naito, S.~Kuwajima, {Simulated low-earth-orbit
  cycle-life testing of commercial laminated lithium-ion cells in a vacuum},
  Journal of Power Sources 140~(1) (2005) 129--138.
\newblock \href {http://dx.doi.org/10.1016/j.jpowsour.2004.08.021}
  {\path{doi:10.1016/j.jpowsour.2004.08.021}}.

\bibitem{Wang2006}
X.~Wang, Y.~Sone, H.~Naito, C.~Yamada, G.~Segami, K.~Kibe, {Cycle-life testing
  of large-capacity lithium-ion cells in simulated satellite operation},
  Journal of Power Sources 161~(1) (2006) 594--600.
\newblock \href {http://dx.doi.org/10.1016/j.jpowsour.2006.04.131}
  {\path{doi:10.1016/j.jpowsour.2006.04.131}}.

\bibitem{Marsh2001}
R.~A. Marsh, S.~Vukson, S.~Surampudi, B.~V. Ratnakumar, M.~C. Smart, M.~Manzo,
  P.~J. Dalton, {Li ion batteries for aerospace applications}, Journal of Power
  Sources 97-98 (2001) 25--27.
\newblock \href {http://dx.doi.org/10.1016/S0378-7753(01)00584-5}
  {\path{doi:10.1016/S0378-7753(01)00584-5}}.

\bibitem{Smart2004}
M.~C. Smart, B.~V. Ratnakumar, L.~D. Whitcanack, K.~B. Chin, S.~Surampudi,
  R.~Gitzendanner, F.~Puglia, J.~Byers, {Batteries for Aerospace}, IEEE
  Aerospace and Electronic Systems Magazine 19~(1) (2004) 18--25.
\newblock \href {http://dx.doi.org/10.1109/MAES.2004.1263988}
  {\path{doi:10.1109/MAES.2004.1263988}}.

\bibitem{Uno2011}
M.~Uno, K.~Ogawa, Y.~Takeda, Y.~Sone, K.~Tanaka, M.~Mita, H.~Saito,
  {Development and on-orbit operation of lithium-ion pouch battery for small
  scientific satellite "'REIMEI"'}, Journal of Power Sources 196~(20) (2011)
  8755--8763.
\newblock \href {http://dx.doi.org/10.1016/j.jpowsour.2011.06.051}
  {\path{doi:10.1016/j.jpowsour.2011.06.051}}.

\bibitem{Sone2016}
Y.~Sone, H.~Watanabe, K.~Tanaka, S.~Fukuda, K.~Ogawa, K.~Asamura, A.~Yamazaki,
  H.~Nagamatsu, Y.~Fukushima, H.~Saito, {Long Term Operability of Li-ion
  Battery under Micro-gravity Condition Demonstrated by the Satellite
  "'REIMEI"'}, Electrochemistry 84~(1) (2016) 12--16.

\bibitem{Saito2011}
H.~Saito, M.~Hirahara, T.~Mizuno, S.~Fukuda, Y.~Fukushima, K.~Asamura,
  H.~Nagamatsu, K.~Tanaka, Y.~Sone, N.~Okuizumi, M.~Mita, M.~Uno, Y.~Yanagawa,
  T.~Takahara, R.~Kaneda, T.~Honma, T.~Sakanoi, A.~Miura, T.~Ikenaga, K.~Ogawa,
  Y.~Masumoto, {Small satellite REIMEI for auroral observations}, Acta
  Astronautica 69~(7-8) (2011) 499--513.
\newblock \href {http://dx.doi.org/10.1016/j.actaastro.2011.05.007}
  {\path{doi:10.1016/j.actaastro.2011.05.007}}.

\bibitem{Brown2008a}
S.~Brown, K.~Ogawa, Y.~Kumeuchi, S.~Enomoto, M.~Uno, H.~Saito, Y.~Sone,
  D.~Abraham, G.~Lindbergh, {Cycle life evaluation of 3 Ah LixMn2O4-based
  lithium-ion secondary cells for low-earth-orbit satellites. I. Full cell
  results}, Journal of Power Sources 185~(2) (2008) 1444--1453.
\newblock \href {http://dx.doi.org/10.1016/j.jpowsour.2008.07.070}
  {\path{doi:10.1016/j.jpowsour.2008.07.070}}.

\bibitem{Attia2019}
P.~M. Attia, S.~Das, S.~J. Harris, M.~Z. Bazant, W.~C. Chueh, {Electrochemical
  kinetics of sei growth on Carbon Black: Part I. experiments}, Journal of the
  Electrochemical Society 166~(4) (2019) E97--E106.
\newblock \href {http://dx.doi.org/10.1149/2.0231904jes}
  {\path{doi:10.1149/2.0231904jes}}.

\bibitem{Boniface2016}
M.~Boniface, L.~Quazuguel, J.~Danet, D.~Guyomard, P.~Moreau,
  P.~Bayle-Guillemaud, {Nanoscale Chemical Evolution of Silicon Negative
  Electrodes Characterized by Low-Loss STEM-EELS}, Nano Letters 16~(12) (2016)
  7381--7388.
\newblock \href {http://dx.doi.org/10.1021/acs.nanolett.6b02883}
  {\path{doi:10.1021/acs.nanolett.6b02883}}.

\bibitem{Huang2019}
W.~Huang, P.~M. Attia, H.~Wang, S.~E. Renfrew, N.~Jin, S.~Das, Z.~Zhang, D.~T.
  Boyle, Y.~Li, M.~Z. Bazant, B.~D. McCloskey, W.~C. Chueh, Y.~Cui, {Evolution
  of the Solid-Electrolyte Interphase on Carbonaceous Anodes Visualized by
  Atomic-Resolution Cryogenic Electron Microscopy}, Nano Letters 19~(8) (2019)
  5140--5148.
\newblock \href {http://dx.doi.org/10.1021/acs.nanolett.9b01515}
  {\path{doi:10.1021/acs.nanolett.9b01515}}.

\bibitem{Rahe2019}
C.~Rahe, S.~T. Kelly, M.~N. Rad, D.~U. Sauer, J.~Mayer, E.~Figgemeier,
  {Nanoscale X-ray imaging of ageing in automotive lithium ion battery cells},
  Journal of Power Sources 433 (2019) 126631.
\newblock \href {http://dx.doi.org/10.1016/j.jpowsour.2019.05.039}
  {\path{doi:10.1016/j.jpowsour.2019.05.039}}.

\bibitem{Pfrang2018}
A.~Pfrang, A.~Kersys, A.~Kriston, D.~U. Sauer, C.~Rahe, S.~K{\"{a}}bitz,
  E.~Figgemeier, {Long-term cycling induced jelly roll deformation in
  commercial 18650 cells}, Journal of Power Sources 392 (2018) 168--175.
\newblock \href {http://dx.doi.org/10.1016/j.jpowsour.2018.03.065}
  {\path{doi:10.1016/j.jpowsour.2018.03.065}}.

\bibitem{Wang2018}
A.~Wang, S.~Kadam, H.~Li, S.~Shi, Y.~Qi, {Review on modeling of the anode solid
  electrolyte interphase (SEI) for lithium-ion batteries}, npj Computational
  Materials 4~(1) (2018) 1--26.
\newblock \href {http://dx.doi.org/10.1038/s41524-018-0064-0}
  {\path{doi:10.1038/s41524-018-0064-0}}.

\bibitem{Bertolini2018}
S.~Bertolini, P.~B. Balbuena, {Buildup of the solid electrolyte interphase on
  lithium-metal anodes: reactive molecular dynamics study}, The Journal of
  Physical Chemistry C 122~(20) (2018) 10783--10791.
\newblock \href {http://dx.doi.org/10.1021/acs.jpcc.8b03046}
  {\path{doi:10.1021/acs.jpcc.8b03046}}.

\bibitem{Schmalstieg2014}
J.~Schmalstieg, S.~K{\"{a}}bitz, M.~Ecker, D.~U. Sauer, {A holistic aging model
  for Li(NiMnCo)O2 based 18650 lithium-ion batteries}, Journal of Power Sources
  257 (2014) 325--334.
\newblock \href {http://dx.doi.org/10.1016/j.jpowsour.2014.02.012}
  {\path{doi:10.1016/j.jpowsour.2014.02.012}}.

\bibitem{Severson2019}
K.~A. Severson, P.~M. Attia, N.~Jin, N.~Perkins, B.~Jiang, Z.~Yang, M.~H. Chen,
  M.~Aykol, P.~K. Herring, D.~Fraggedakis, M.~Z. Bazant, S.~J. Harris, W.~C.
  Chueh, R.~D. Braatz, {Data-driven prediction of battery cycle life before
  capacity degradation}, Nature Energy 4~(5) (2019) 383--391.
\newblock \href {http://dx.doi.org/10.1038/s41560-019-0356-8}
  {\path{doi:10.1038/s41560-019-0356-8}}.

\bibitem{Roder2017}
F.~R{\"{o}}der, R.~D. Braatz, U.~Krewer, {Multi-scale simulation of
  heterogeneous surface film growth mechanisms in lithium-ion batteries},
  Journal of the Electrochemical Society 164~(11) (2017) E3335--E3344.
\newblock \href {http://dx.doi.org/10.1149/2.0241711jes}
  {\path{doi:10.1149/2.0241711jes}}.

\bibitem{Das2019}
S.~Das, P.~M. Attia, W.~C. Chueh, M.~Z. Bazant, {Electrochemical kinetics of
  sei growth on carbon black: Part II. Modeling}, Journal of the
  Electrochemical Society 166~(4) (2019) E107--E118.
\newblock \href {http://dx.doi.org/10.1149/2.0241904jes}
  {\path{doi:10.1149/2.0241904jes}}.

\bibitem{Lin2016}
Y.~X. Lin, Z.~Liu, K.~Leung, L.~Q. Chen, P.~Lu, Y.~Qi, {Connecting the
  irreversible capacity loss in Li-ion batteries with the electronic insulating
  properties of solid electrolyte interphase (SEI) components}, Journal of
  Power Sources 309 (2016) 221--230.
\newblock \href {http://dx.doi.org/10.1016/j.jpowsour.2016.01.078}
  {\path{doi:10.1016/j.jpowsour.2016.01.078}}.

\bibitem{Li2015}
D.~Li, D.~Danilov, Z.~Zhang, H.~Chen, Y.~Yang, P.~H.~L. Notten, {Modeling the
  SEI-Formation on Graphite Electrodes in LiFePO4 Batteries}, Journal of The
  Electrochemical Society 162~(6) (2015) A858--A869.
\newblock \href {http://dx.doi.org/10.1149/2.0161506jes}
  {\path{doi:10.1149/2.0161506jes}}.

\bibitem{Heinrich2019}
M.~Heinrich, N.~Wolff, N.~Harting, V.~Laue, F.~R{\"{o}}der, S.~Seitz,
  U.~Krewer, {Physico-Chemical Modeling of a Lithium-Ion Battery: An Ageing
  Study with Electrochemical Impedance Spectroscopy}, Batteries and Supercaps
  2~(6) (2019) 530--540.
\newblock \href {http://dx.doi.org/10.1002/batt.201900011}
  {\path{doi:10.1002/batt.201900011}}.

\bibitem{Chouchane2021}
M.~Chouchane, O.~Arcelus, A.~A. Franco, {Heterogeneous Solid-Electrolyte
  Interphase in Graphite Electrodes Assessed by 4D-Resolved Computational
  Simulations}, Batteries {\&} Supercaps 4.
\newblock \href {http://dx.doi.org/10.1002/batt.202100030}
  {\path{doi:10.1002/batt.202100030}}.

\bibitem{Single2018}
F.~Single, A.~Latz, B.~Horstmann, {Identifying the Mechanism of Continued
  Growth of the Solid-Electrolyte Interphase}, ChemSusChem 11~(12) (2018)
  1950--1955.
\newblock \href {http://dx.doi.org/10.1002/cssc.201800077}
  {\path{doi:10.1002/cssc.201800077}}.

\bibitem{Soto2015}
F.~A. Soto, Y.~Ma, J.~M. {Martinez De La Hoz}, J.~M. Seminario, P.~B. Balbuena,
  {Formation and Growth Mechanisms of Solid-Electrolyte Interphase Layers in
  Rechargeable Batteries}, Chemistry of Materials 27~(23) (2015) 7990--8000.
\newblock \href {http://dx.doi.org/10.1021/acs.chemmater.5b03358}
  {\path{doi:10.1021/acs.chemmater.5b03358}}.

\bibitem{Pinson2013}
M.~B. Pinson, M.~Z. Bazant, {Theory of SEI formation in rechargeable batteries:
  Capacity fade, accelerated aging and lifetime prediction}, Journal of the
  Electrochemical Society 160~(2) (2013) A243--A250.
\newblock \href {http://dx.doi.org/10.1149/2.044302jes}
  {\path{doi:10.1149/2.044302jes}}.

\bibitem{Ploehn2004}
H.~J. Ploehn, P.~Ramadass, R.~E. White, {Solvent Diffusion Model for Aging of
  Lithium-Ion Battery Cells}, Journal of The Electrochemical Society 151~(3)
  (2004) A456--A462.
\newblock \href {http://dx.doi.org/10.1149/1.1644601}
  {\path{doi:10.1149/1.1644601}}.

\bibitem{Ekstrom2015}
H.~Ekstr{\"{o}}m, G.~Lindbergh, { A Model for Predicting Capacity Fade due to
  SEI Formation in a Commercial Graphite/LiFePO 4 Cell }, Journal of The
  Electrochemical Society 162~(6) (2015) A1003--A1007.
\newblock \href {http://dx.doi.org/10.1149/2.0641506jes}
  {\path{doi:10.1149/2.0641506jes}}.

\bibitem{Roder2016}
F.~R{\"{o}}der, R.~D. Braatz, U.~Krewer, {Multi-Scale Modeling of Solid
  Electrolyte Interface Formation in Lithium-Ion Batteries}, in: Computer Aided
  Chemical Engineering, Vol.~38, Elsevier, 2016, pp. 157--162.
\newblock \href {http://dx.doi.org/10.1016/B978-0-444-63428-3.50031-X}
  {\path{doi:10.1016/B978-0-444-63428-3.50031-X}}.

\bibitem{VonKolzenberg2020}
L.~von Kolzenberg, A.~Latz, B.~Horstmann, {Solid–Electrolyte Interphase
  During Battery Cycling: Theory of Growth Regimes}, ChemSusChem 13~(15) (2020)
  3901--3910.
\newblock \href {http://arxiv.org/abs/2004.01458} {\path{arXiv:2004.01458}},
  \href {http://dx.doi.org/10.1002/cssc.202000867}
  {\path{doi:10.1002/cssc.202000867}}.

\bibitem{Horstmann2019}
B.~Horstmann, F.~Single, A.~Latz, {Review on multi-scale models of
  solid-electrolyte interphase formation}, Current Opinion in Electrochemistry
  13 (2019) 61--69.
\newblock \href {http://dx.doi.org/10.1016/j.coelec.2018.10.013}
  {\path{doi:10.1016/j.coelec.2018.10.013}}.

\bibitem{Tahmasbi_2017}
A.~A. Tahmasbi, T.~Kadyk, M.~H. Eikerling, Statistical physics-based model of
  solid electrolyte interphase growth in lithium ion batteries, Journal of The
  Electrochemical Society 164~(6) (2017) A1307--A1313.
\newblock \href {http://dx.doi.org/10.1149/2.1581706jes}
  {\path{doi:10.1149/2.1581706jes}}.

\bibitem{BEST}
ITWM, {BEST - Battery and Electrochemistry Simulation Tool},
  \url{http://itwm.fraunhofer.de/best} (2020).

\bibitem{Schmitt2020}
T.~Schmitt, Degradation models and simulation tools for lithium and zinc
  batteries, Ph.D. thesis, Universität Ulm (2020).

\bibitem{Latz2011}
A.~Latz, J.~Zausch, {Thermodynamic consistent transport theory of Li-ion
  batteries}, Journal of Power Sources 196~(6) (2011) 3296--3302.
\newblock \href {http://dx.doi.org/10.1016/j.jpowsour.2010.11.088}
  {\path{doi:10.1016/j.jpowsour.2010.11.088}}.

\bibitem{Mendoza2020}
O.~S. Mendoza-Hernandez, L.~J. Bolay, B.~Horstmann, A.~Latz, E.~Hosono,
  D.~Asakura, H.~Matsuda, M.~Itagaki, M.~Umeda, Y.~Sone, Durability analysis of
  the reimei satellite li-ion batteries after more than 14 years of operation
  in space, Electrochemistry 88~(4) (2020) 300--304.
\newblock \href {http://dx.doi.org/10.5796/electrochemistry.20-00046}
  {\path{doi:10.5796/electrochemistry.20-00046}}.

\bibitem{Forman2012}
J.~C. Forman, S.~J. Moura, J.~L. Stein, H.~K. Fathy, {Genetic identification
  and fisher identifiability analysis of the Doyle-Fuller-Newman model from
  experimental cycling of a LiFePO 4 cell}, Journal of Power Sources 210 (2012)
  263--275.
\newblock \href {http://dx.doi.org/10.1016/j.jpowsour.2012.03.009}
  {\path{doi:10.1016/j.jpowsour.2012.03.009}}.

\bibitem{Rahman2016}
A.~Rahman, S.~Anwar, A.~Izadian, {Electrochemical model parameter
  identification of a lithium-ion battery using particle swarm optimization
  method}, Journal of Power Sources 307 (2016) 86--97.
\newblock \href {http://dx.doi.org/10.1016/j.jpowsour.2015.12.083}
  {\path{doi:10.1016/j.jpowsour.2015.12.083}}.

\bibitem{Sone2017}
Y.~Sone, H.~Watanabe, K.~Tanaka, O.~Mendoza-Hernandez, S.~Fukuda, M.~Itagaki,
  K.~Ogawa, K.~Asamura, A.~Yamazaki, H.~Nagamatsu, Y.~Fukushima, H.~Saito,
  {Internal Impedance of the Lithium-Ion Secondary Cells Used for Reimei
  Satellite after the Eleven Years Operation in Space}, E3S Web of Conferences
  16~(3) (2017) 4--8.
\newblock \href {http://dx.doi.org/10.1051/e3sconf/20171607005}
  {\path{doi:10.1051/e3sconf/20171607005}}.

\bibitem{Brown2008b}
S.~Brown, K.~Ogawa, Y.~Kumeuchi, S.~Enomoto, M.~Uno, H.~Saito, Y.~Sone,
  D.~Abraham, G.~Lindbergh, {Cycle life evaluation of 3 Ah LixMn2O4-based
  lithium-ion secondary cells for low-earth-orbit satellites. II. Harvested
  electrode examination}, Journal of Power Sources 185~(2) (2008) 1454--1464.
\newblock \href {http://dx.doi.org/10.1016/j.jpowsour.2008.07.070}
  {\path{doi:10.1016/j.jpowsour.2008.07.070}}.

\bibitem{Prifling2019}
B.~Prifling, A.~Ridder, A.~Hilger, M.~Osenberg, I.~Manke, K.~P. Birke,
  V.~Schmidt, {Analysis of structural and functional aging of electrodes in
  lithium-ion batteries during rapid charge and discharge rates using
  synchrotron tomography}, Journal of Power Sources 443 (2019) 227259.
\newblock \href {http://dx.doi.org/10.1016/j.jpowsour.2019.227259}
  {\path{doi:10.1016/j.jpowsour.2019.227259}}.

\bibitem{geodict}
Math2Market GmbH, Deutschland, {GeoDict 2019},
  \url{https://www.math2market.com/}.

\bibitem{Hein2016a}
S.~Hein, J.~Feinauer, D.~Westhoff, I.~Manke, V.~Schmidt, A.~Latz, {Stochastic
  microstructure modeling and electrochemical simulation of lithium-ion cell
  anodes in 3D}, Journal of Power Sources 336 (2016) 161--171.
\newblock \href {http://dx.doi.org/10.1016/j.jpowsour.2016.10.057}
  {\path{doi:10.1016/j.jpowsour.2016.10.057}}.

\bibitem{Latz2015}
A.~Latz, J.~Zausch, {Multiscale modeling of lithium ion batteries: Thermal
  aspects}, Beilstein Journal of Nanotechnology 6~(1) (2015) 987--1007.
\newblock \href {http://dx.doi.org/10.3762/bjnano.6.102}
  {\path{doi:10.3762/bjnano.6.102}}.

\bibitem{Hein2016}
S.~Hein, A.~Latz, {Influence of local lithium metal deposition in 3D
  microstructures on local and global behavior of Lithium-ion batteries},
  Electrochimica Acta 201 (2016) 354--365.
\newblock \href {http://dx.doi.org/10.1016/j.electacta.2016.01.220}
  {\path{doi:10.1016/j.electacta.2016.01.220}}.

\bibitem{Schmitt2020a}
T.~Schmitt, A.~Latz, B.~Horstmann, {Derivation of a local volume-averaged model
  and a stable numerical algorithm for multi-dimensional simulations of
  conversion batteries}, Electrochimica Acta 333 (2020) 135491.
\newblock \href {http://dx.doi.org/10.1016/J.ELECTACTA.2019.135491}
  {\path{doi:10.1016/J.ELECTACTA.2019.135491}}.

\bibitem{Doyle1993}
M.~Doyle, T.~F. Fuller, J.~Newman, {Modeling of Galvanostatic Charge and
  Discharge of the Lithium/Polymer/Insertion Cell}, Journal of The
  Electrochemical Society 140~(6) (1993) 1526--1533.
\newblock \href {http://dx.doi.org/10.1149/1.2221597}
  {\path{doi:10.1149/1.2221597}}.

\bibitem{Bizeray2015}
A.~M. Bizeray, S.~Zhao, S.~R. Duncan, D.~A. Howey, {Lithium-ion battery
  thermal-electrochemical model-based state estimation using orthogonal
  collocation and a modified extended Kalman filter}, Journal of Power Sources
  296 (2015) 400--412.
\newblock \href {http://arxiv.org/abs/1506.08689} {\path{arXiv:1506.08689}},
  \href {http://dx.doi.org/10.1016/j.jpowsour.2015.07.019}
  {\path{doi:10.1016/j.jpowsour.2015.07.019}}.

\bibitem{doi:10.1021/ja305366r}
S.~Shi, P.~Lu, Z.~Liu, Y.~Qi, L.~G. Hector, H.~Li, S.~J. Harris, Direct
  calculation of li-ion transport in the solid electrolyte interphase, Journal
  of the American Chemical Society 134~(37) (2012) 15476--15487.
\newblock \href {http://dx.doi.org/10.1021/ja305366r}
  {\path{doi:10.1021/ja305366r}}.

\bibitem{Ehrl2017}
A.~Ehrl, J.~Landesfeind, W.~A. Wall, H.~A. Gasteiger, {Determination of
  Transport Parameters in Liquid Binary Lithium Ion Battery Electrolytes},
  Journal of The Electrochemical Society 164~(4) (2017) A826--A836.
\newblock \href {http://dx.doi.org/10.1149/2.1131704jes}
  {\path{doi:10.1149/2.1131704jes}}.

\bibitem{Regis2007}
R.~G. Regis, C.~A. Shoemaker, {A Stochastic Radial Basis Function Method for
  the Global Optimization of Expensive Functions}, INFORMS Journal on Computing
  19~(4) (2007) 497--509.
\newblock \href {http://dx.doi.org/10.1287/ijoc.1060.0182}
  {\path{doi:10.1287/ijoc.1060.0182}}.

\bibitem{Wang2014}
Y.~Wang, C.~A. Shoemaker, {A General Stochastic Algorithmic Framework for
  Minimizing Expensive Black Box Objective Functions Based on Surrogate Models
  and Sensitivity Analysis}, arXiv preprint arXiv:1410.6271 (2014) .

\bibitem{Bizeray2018}
A.~M. Bizeray, J.~H. Kim, S.~R. Duncan, D.~A. Howey, {Identifiability and
  Parameter Estimation of the Single Particle Lithium-Ion Battery Model}, IEEE
  Transactions on Control Systems Technology 27~(5) (2019) 1862--1877.
\newblock \href {http://dx.doi.org/10.1109/TCST.2018.2838097}
  {\path{doi:10.1109/TCST.2018.2838097}}.

\bibitem{Li2020}
W.~Li, D.~Cao, D.~J{\"{o}}st, F.~Ringbeck, M.~Kuipers, F.~Frie, D.~U. Sauer,
  {Parameter sensitivity analysis of electrochemical model-based battery
  management systems for lithium-ion batteries}, Applied Energy 269 (2020)
  115104.
\newblock \href {http://dx.doi.org/10.1016/j.apenergy.2020.115104}
  {\path{doi:10.1016/j.apenergy.2020.115104}}.

\bibitem{MATLAB}
The MathWorks, Inc., Natick, Massachusetts, {MATLAB 2020a},
  \url{https://www.mathworks.com/products/matlab.html}.

\bibitem{Reniers2019}
J.~M. Reniers, G.~Mulder, D.~A. Howey, {Review and Performance Comparison of
  Mechanical-Chemical Degradation Models for Lithium-Ion Batteries}, Journal of
  The Electrochemical Society 166~(14) (2019) A3189--A3200.
\newblock \href {http://dx.doi.org/10.1149/2.0281914jes}
  {\path{doi:10.1149/2.0281914jes}}.

\bibitem{Shi2015}
S.~Shi, J.~Gao, Y.~Liu, Y.~Zhao, Q.~Wu, W.~Ju, C.~Ouyang, R.~Xiao, {Multi-scale
  computation methods: Their applications in lithium-ion battery research and
  development}, Chinese Physics B 25~(1) (2016) 018212.
\newblock \href {http://dx.doi.org/10.1088/1674-1056/25/1/018212}
  {\path{doi:10.1088/1674-1056/25/1/018212}}.

\end{thebibliography}

\end{document}